\documentstyle[preprint,aps]{revtex}
\begin{document}
\preprint{
\begin{minipage}[t]{1.8in}
NHMFL-93-988\\[-0.8em]
ETH-TH/93-33
\end{minipage}
}
\draft
\title{Evidence for Spin-Charge Separation in the\\
Two-Dimensional $t$-$J$ Model}
\author{W. O. Putikka}
\address{National High Magnetic Field Laboratory,
Florida State University, Tallahassee, FL 32306}
\author{R. L. Glenister and R. R. P. Singh}
\address{Department of Physics, University of California,
Davis, California 95616}
\author{Hirokazu Tsunetsugu}
\address{Theoretische Physik, Eidgen\"ossische Technische Hochschule,
8093 Z\"urich, Switzerland and\\
Interdisziplin\"ares Projektzentrum f\"ur Supercomputing,
ETH-Zentrum, 8092 Z\"urich, Switzerland}
\date{September 20, 1993}
\maketitle
\begin{abstract}
We have calculated high temperature expansions for the momentum
distribution function $n_{\scriptscriptstyle\bf k}$ and the equal time
spin and density correlation functions $S({\bf q})$ and $N({\bf q})$ of
the two-dimensional $t$-$J$ model. On extrapolation to low temperatures
we find that $n_{\scriptscriptstyle\bf k}$ has a step-like feature at
${\bf k}_{\scriptscriptstyle F}$ and $S({\bf q})$ has $2{\bf
k}_{\scriptscriptstyle F}$ as a characteristic wavevector, whereas
$N({\bf q})$ has $2{\bf k}_{\scriptscriptstyle F}^{\scriptscriptstyle
SF}$ as a characteristic wavevector.  Here ${\bf k}_{\scriptscriptstyle
F}$ and ${\bf k}_{\scriptscriptstyle F}^{\scriptscriptstyle SF}$ are the
Fermi wavevectors of the nearest-neighbor square lattice tight-binding
and spinless fermion models, respectively. By comparison to the known
results for one dimension this suggests spin-charge separation in the
two-dimensional $t$-$J$ model.
\end{abstract}
The study of strongly correlated electrons in two-dimensions (2D) is
currently one of the most interesting and controversial topics in
condensed matter physics, particularly with regard to high temperature
superconductivity (HTSC)\cite{TMR}.  Anderson\cite{PWA1} has put forward
the idea that the ground state of strongly correlated electron systems
in 2D is a Luttinger liquid analogous to the case in one dimension.  In
1D, a Luttinger liquid has spin and charge degrees of freedom with
different velocities and wavevectors, behaving at low energies as
independent elementary excitations, a situation which has become known
as spin-charge separation\cite{PWA1}.

Determining whether or not spin-charge separation can also occur in 2D
has proven quite difficult.  We have calculated high temperature
expansions for equal time correlation functions (ETCF) of the 2D $t$-$J$
model\cite{ZR} to investigate this possibility.  We find two distinct
characteristic wavevectors for the spin and charge degrees of freedom,
$2{\bf k}_{\scriptscriptstyle F}$ and $2{\bf k}_{\scriptscriptstyle
F}^{\scriptscriptstyle SF}$ defined below.  This shows that the spin and
charge degrees of freedom have different distributions in the Brillouin
zone and provides evidence for spin-charge separation in this model.

We consider the 2D $t$-$J$ model on a square lattice, where the
Hamiltonian is
\begin{equation}
{\cal H}_{tJ}=-t\sum_{<ij>,\sigma} ( c_{i\sigma}^\dagger c_{j\sigma}
+ h.c.) + J\sum_{<ij>} {\bf S}_i\cdot{\bf S}_j,
\end{equation}
with the constraint of no double occupancy. The constraint represents
the strong correlations between the electrons and is difficult to treat
by conventional many-body techniques.

We have studied three ETCF of this model using the high temperature
series expansion method.  These are the single spin momentum
distribution function, $n_{\scriptscriptstyle\bf k}$, and the equal time
spin and density correlation functions, $S({\bf q})$ and $N({\bf q})$,
defined by the relations
\begin{eqnarray}
n_{\scriptscriptstyle\bf k}&=&
\sum_{{\bf r}}e^{i{\bf k\cdot r}}\langle
c_{0\sigma}^{\dagger}c_{{\bf r}\sigma}\rangle,\nonumber\\
S({\bf q})&=&\sum_{\bf r}e^{i{\bf q\cdot r}}\langle
S^{z}_{0}S^{z}_{\bf r}\rangle,\nonumber\\
N({\bf q})&=&\sum_{\bf r}e^{i{\bf q\cdot r}}\langle
\Delta n_{0}\Delta n_{\bf r}\rangle,
\end{eqnarray}
where the angular brackets refer to thermal averaging in the grand
canonical ensemble, $S^{z}_{\bf r}={1\over2}
\sum_{\alpha\beta}c^{\dagger}_{{\bf r}\alpha}\sigma
^{z}_{\alpha\beta}c_{{\bf r}\beta}$ and $\Delta n_{\bf r}=\sum_{\sigma}
c^{\dagger}_{{\bf r}\sigma}c_{{\bf r}\sigma}-n$.  Here $n$ is the
average density of electrons.  The expansions are calculated for
$n_{\scriptscriptstyle\bf k}$\cite{SG1} to eighth order and $S({\bf
q})$\cite{SG2} and $N({\bf q})$, to tenth order in the reciprocal
temperature $T^{-1}$, first at fixed fugacity, and then by a change of
variables at fixed $n$.  The series are extrapolated at fixed $n$ by
Pad\'e approximants to determine the low $T$ properties.

The form of the ETCF for tight-binding (TB) (non-interacting, spin-half
electrons with nearest neighbor hopping on a square lattice) and
spinless fermions (SF) (physically, SF are fully spin-polarized TB
electrons, freezing out the spin degrees of freedom and doubling the
number of occupied ${\bf k}$ states) in 2D are helpful in understanding
the $t$-$J$ model results presented below.  They are given by
\begin{equation}
 N({\bf q}) = n-g\int{d{\bf k}\over(2\pi)^2}n_{\scriptscriptstyle\bf k}
n_{\scriptscriptstyle\bf k+q},
\end{equation}
where $g=2$ for TB or $g=1$ for SF, and $4S({\bf q})=N({\bf q})$ for TB.
{}From the form of this equation we can see that at $T=0$ and $n\le1$
the ETCF will saturate at $n$ when $n_{\scriptscriptstyle\bf k}$ and
$n_{\scriptscriptstyle\bf k+q}$ no longer overlap (note that for SF with
$n>0.5$, $N({\bf q})$ saturates at $1-n$ when the hole Fermi surfaces no
longer overlap).  The Kohn anomaly\cite{WK} at $2{\bf
k}_{\scriptscriptstyle F}$ or $2{\bf k}_{\scriptscriptstyle
F}^{\scriptscriptstyle SF}$ is due to the existence of a sharp Fermi
surface.

In Fig.\ \ref{fig1} we compare $N({\bf q})$ of the 2D $t$-$J$ model to
$N({\bf q})$ of SF at the same density for $T/J=0.5$, $J/t=0.5$ and a
range of $n$.  The similarities are remarkable throughout the Brillouin
zone, with the differences near the $\Gamma$ point due to the $t$-$J$
model having a larger compressibility than SF.  To focus the discussion
below we now limit ourselves to two sets of parameters outside of the
phase separated\cite{PLR} or ferromagnetic\cite{PLO} regions of the 2D
$t$-$J$ model.  We fix $J/t=0.5$ and consider $n=0.75$ and $n=0.20$. The
results for $n_{\scriptscriptstyle\bf k}$, $S({\bf q})$ and $N({\bf q})$
along $q_{\scriptscriptstyle\Gamma M}=(0,0)\to(\pi,\pi)$ are shown in
Fig.\ \ref{fig2}.  We see that $n_{\scriptscriptstyle \bf k}\approx 1/2$
at $k_{\scriptscriptstyle F\ \Gamma M}$, the Fermi momentum of the TB
model at the same density\cite{SG1,SH} and $S({\bf q})$ is enhanced over
its TB value and either flattens out or has a peak\cite{SG2} at
$q\approx 2 k_{\scriptscriptstyle F\ \Gamma M}$.  However, the most
anomalous curves are for $N({\bf q})$.  They are suppressed from their
TB values and flatten out at $q\approx 2 k_{\scriptscriptstyle F\ \Gamma
M}^{\scriptscriptstyle SF}$, the Fermi momentum of SF at the same
density.  We observe no feature at ${\bf q}=2{\bf k}_{\scriptscriptstyle
F}$, though $N({\bf q})$ flattens out more gradually for $n=0.20$ than
for $n=0.75$.

For comparison, we recall the behaviors of $n_{\scriptscriptstyle k}$,
$S(q)$ and $N(q)$ for the $U/t\to\infty$ Hubbard model ($J/t\to 0$
$t$-$J$ model) in 1D. From the work of Ogata and Shiba\cite{OS} we know
that $n_{\scriptscriptstyle k}$ has a power law singularity at
$k_{\scriptscriptstyle F}$ with $n_{k_F}=1/2$.  Also $S(q)$ has a peak
at $2k_{\scriptscriptstyle F}$ and $N(q)$ has $2k_{\scriptscriptstyle
F}^{\scriptscriptstyle SF} = 4k_{\scriptscriptstyle F}$ as a
characteristic wavevector.  For arbitrary $U/t$, $N(q)$ also has a
feature at $2k_{\scriptscriptstyle F}$ due to a mixture of spin and
charge excitations, but the $2k_{\scriptscriptstyle
F}^{\scriptscriptstyle SF}$ feature is due to charge alone\cite{FK} and
shows that the charge degrees of freedom truly reside at
$k_{\scriptscriptstyle F}^{\scriptscriptstyle SF}$.  In addition to the
singularity at $k_{\scriptscriptstyle F}$, $n_{\scriptscriptstyle k}$
has a singularity at $3k_{\scriptscriptstyle F}$, but with a very small
step size.

Our calculated 2D ETCF show behaviors very similar to their counterparts
in 1D.  The characteristic wavevectors for $S({\bf q})$ and $N({\bf q})$
are $2{\bf k}_{\scriptscriptstyle F}$ and $2{\bf k}_{\scriptscriptstyle
F}^{\scriptscriptstyle SF}$, respectively, which we believe implies low
energy spin degrees of freedom near ${\bf k}_{\scriptscriptstyle F}$ and
low energy charge degrees of freedom near ${\bf k}_{\scriptscriptstyle
F}^{\scriptscriptstyle SF}$.  Note that in 2D ${\bf
k}_{\scriptscriptstyle F}$ and ${\bf k}_{\scriptscriptstyle
F}^{\scriptscriptstyle SF}$ are {\it incommensurate} wavevectors; the
charge degrees of freedom do not occur at a harmonic of ${\bf
k}_{\scriptscriptstyle F}$, but at an independent wavevector.  In Fig.\
\ref{fig4} we show ${\bf k}_{\scriptscriptstyle F}$ and ${\bf
k}_{\scriptscriptstyle F}^{\scriptscriptstyle SF}$ for the whole
Brillouin zone at $n=0.75$ with representative nesting wavevectors.  For
weak coupling calculations of the 2D Hubbard model\cite{SLH} and
Gutzwiller projected free electrons\cite{YS} the picture is quite
different.  In these cases while $S({\bf q})$ is enhanced and $N({\bf
q})$ is supressed, they both have  $2{\bf
k}_{\scriptscriptstyle F}$ as a characteristic wavevector
which is not what we find for the 2D $t$-$J$
model.  The behavior of $n_{\scriptscriptstyle\bf k}$ is also similar to
$1D$.  The step in $n_{\scriptscriptstyle\bf k}$ at $n=0.20$ is
comparable in size and shape to the TB model at the same $T$, but at
$n=0.75$ the step is much weaker and too smeared out to be explained by
thermal broadening alone\cite{SG1}.  We have not seen any evidence for a
singularity at $3{\bf k}_{\scriptscriptstyle F}$ in 2D. This could be
due to the relatively high temperature $T/J=1.0$ in our calculation or
possibly the angular averaging in 2D which is not present in 1D.

In 1D the statistics of the excitations play no role, but in 2D they are
important. Our data give no direct evidence on the statistics of the
excitations in 2D, but we can formulate a hypothesis as to what they
might be\cite{PWA1}. If we think of a single electron as being composed
of an elementary spin degree of freedom and an elementary charge degree
of freedom, we would expect one of them to be fermionic and the other
bosonic to give a fermionic electron\cite{Anyon}.  Since
$n_{\scriptscriptstyle\bf k}$ shows a step at ${\bf
k}_{\scriptscriptstyle F}$ and $S({\bf q})$ has $2{\bf
k}_{\scriptscriptstyle F}$ as a characteristic wavevector we assign the
spin degrees of freedom as fermionic and the charge degrees of freedom
as bosonic, but note that the charge degrees of freedom are not free
bosons, but hard core bosons (HCB) to enforce the constraint of no
double occupancy.  This can be seen in Fig.\ \ref{fig1} for $n=0.5$
where the data points near $(\pi,
\pi)$ are already more rounded than SF at $T/J=0.5$.  Further evidence
for this point of view can be gained from the work of Long and
Zotos\cite{LZ} and Sorella, Parola and Tosatti\cite{SPT}.

We have also estimated the behavior of the HCB $N({\bf q})$ by a flux
phase mean field calculation.  In 2D, HCB on a square lattice with
nearest neighbor hopping can be exactly mapped into SF by attaching a
quantum of magnetic flux $\phi_0$ to each particle\cite{FW}. If density
fluctuations are not large we can replace the attached flux tubes by a
uniform magnetic field, $B_0 = n
\phi_0$, which will couple to the orbital motion of the particles. This
corresponds to a SF model with a site dependent phase (the uniform flux
phase).  Using this flux phase mean field approximation we have
calculated $N({\bf q})$, with the results at $T=0$ shown in Fig. 2.  The
global features show a rounded flattening out of $N({\bf q})$ at $2{\bf
k}_{\scriptscriptstyle F}^{\scriptscriptstyle SF}$ and general agreement
with SF and the $t$-$J$ model.  For small ${\bf q}$ the approximation we
are using gives $N({\bf q})\propto q^2$, but by general hydrodynamic
arguments we know that for $T=0$ if the system has a finite, non-zero
compressibility the $q \rightarrow 0$ limit should be linear.  The
quadratic dependence is due to the ``Fermi energy'' of the flux phase
sitting in an energy gap\cite{HH}.  Therefore the $q^2$-dependence is an
artifact due to our mean field approximation and should become linear
after including fluctuations, which we will discuss in a future paper.

More information on the interactions between the spin and charge degrees
of freedom could be obtained by considering the 2D $t$-$J$ model with a
non-zero spin polarization.  If the spin and charge are coupled we would
expect both $S({\bf q})$ and $N({\bf q})$ to change.  However, if the
spin and charge degrees of freedom are truly separate the characteristic
${\bf q}$-vector of $N({\bf q})$ should not be affected by a non-zero
spin polarization\cite{SF2} but $S({\bf q})$ would now have transverse
and longitudinal components with features at wavevectors that depend on
the number of up and down spins.  This has been observed by Ogata,
Sugiyama and Shiba\cite{OSS} for the 1D $U\to\infty$ Hubbard model.

Having elementary degrees of freedom at ${\bf k}_{\scriptscriptstyle F}$
and ${\bf k}_{\scriptscriptstyle F}^{\scriptscriptstyle SF}$ should have
experimental consequences for the copper oxide planes in HTSC.  Neutron
scattering experiments\cite{MAM} on La$_{2-x}$Sr$_x$CuO$_4$ show four
incommensurate peaks centered around $(\pi,\pi)$ that move with doping.
This can be understood in terms of the nesting properties of the weak
coupling Fermi surface\cite{LZAM} where our results also put the spin
degrees of freedom.  The energy integrated weight of angle resolved
photoemission is a direct measure of $n_{\scriptscriptstyle\bf k}$ and
photoemission has also been interpreted as supporting a large Fermi
surface\cite{CGO}.  But the transport measurements\cite{GLA} are not so
easy to understand in this picture.

Reconciling experiments which show a large electron-like contour in
${\bf k}$-space with a density of $n$ carriers, with transport
measurements, which show a much smaller hole density of $1-n$ carriers,
is one of the most puzzling problems of the copper oxides\cite{GLA}.
Our results for the 2D $t$-$J$ model show one way this might
occur\cite{PWA1,FH}. One expects the transport experiments to couple
most strongly to charge.  For $1-n\ll 1$, the charge degrees of freedom
at ${\bf k}_{\scriptscriptstyle F}^{\scriptscriptstyle SF}$ have a small
{\it hole-like} locus in momentum space centered around $(\pi,\pi)$ as
shown in Fig.\ \ref{fig4}.  With this the transport measurements are
satisfied.  At the same time the spin degrees of freedom give a large,
weak coupling Fermi surface also shown in Fig.\ \ref{fig4}, which is
seen in neutron scattering and photoemission experiments.  We wish to
emphasize that experiments which could probe $N({\bf q})$ directly may
prove to be very interesting for HTSC materials.

In conclusion, we have studied the equal time correlation functions of
the 2D $t$-$J$ model by high temperature series expansion methods.  We
find that the spin and charge ETCF exhibit signatures of two different
wavevectors: the characteristic wavevector for the spins being ${\bf
k}_{\scriptscriptstyle F}$ and that for charge ${\bf
k}_{\scriptscriptstyle F}^{\scriptscriptstyle SF}$, the Fermi
wavevectors for TB and SF respectively. In comparison with the results
for 1D this suggests spin-charge separation in this strongly correlated
2D model.

WOP was supported by NSF Grant No. DMR-91-14553 and by the National High
Magnetic Field Laboratory at Florida State University.  RLG and RRPS
were supported by NSF Grant No. DMR-9017361.  HT was supported by the
Swiss National Science Foundation Grant No. NFP-304030-032833.  The
authors thank T. M. Rice for many useful conversations and a critical
reading of the manuscript.  The authors also thank N. E. Bonesteel, H.
Fukuyama, M. Ogata, N. P. Ong, D. J. Scalapino, J. R. Schrieffer and G.
Zimanyi for many useful discussions.  WOP and RRPS thank the CMS group
at Los Alamos for hospitality while this manuscript was being completed.
Part of the computations were done on a Cray YMP at Cray Research, Inc.

{\it Note added in proof}--After completion of this work we received a
preprint by Y. C. Chen and T. K. Lee \cite{CL} which arrives at results
similar to ours.

\begin{figure}
\caption{
Plot of $N({\bf q})$ at $T/J=0.5$ and $J/t=0.5$ along the irreducible
wedge for a range of $n$.  The data points are the $t$-$J$ model and the
solid lines are spinless fermions for the same temperature.  The small
vertical arrows are the $T=0$ locations of nesting vectors for spinless
fermions.}
\label{fig1}
\end{figure}

\begin{figure}
\caption{
Plots along the diagonal $\Gamma \rightarrow M$ at $n=0.20$ (a) Single
spin momentum distribution function.  Data points: $t$-$J$ model, solid
line: tight-binding model at $T/J=1.0$; (b) Spin correlation function.
Data points: $t$-$J$ model, dashed line: $T=0$ tight-binding model; (c)
Density correlation function.  Data points: $t$-$J$ model, solid line:
$T=0$ flux phase mean field approximation for hard core bosons, dashed
line: $T=0$ spinless fermions and dotted line: $T=0$ tight-binding
model.  The vertical dashed lines indicate the important wavevectors
along this line in the Brillouin zone for tight-binding electrons and
spinless fermions: nesting wavevectors $2k_{\scriptscriptstyle F\ \Gamma
M}$ and $2k_{\scriptscriptstyle F\
\Gamma M}^{\scriptscriptstyle SF}$, or Fermi wavevectors
$k_{\scriptscriptstyle F\ \Gamma M}$ and $k_{\scriptscriptstyle F\
\Gamma M}^{\scriptscriptstyle SF}$.  (d) - (f) same as (a) - (c) with
$n=0.75$.  $K_{\scriptscriptstyle \Gamma M}=(2\pi,2\pi)$ is a reciprocal
lattice vector.}
\label{fig2}
\end{figure}

\begin{figure}
\caption{
Fermi wavevectors for $n=0.75$. Solid curve: tight-binding electrons,
dashed curve: spinless fermions.  The arrows are representative nesting
wavevectors along $\Gamma M$ and $MX$.}
\label{fig4}
\end{figure}


\begin{references}
\bibitem{Anyon} We do not consider the possibility of anyons, which
cannot be ruled out.
\end{references}
\end{document}